# Improvement of the guiding performances of near infrared organic/inorganic channel waveguides.

M. Oubaha[a,b,c], P. Etienne[a*], P. Boutinaud[b], J. M. Nedelec[b], P. Coudray[c] and Y. Moreau[c]

[a] Laboratoire Des Verres, UMR 5587, Université de Montpellier II, Place Eugène Bataillon, Montpellier cedex 5 34090, France

[b] Laboratoire des Matériaux Inorganiques CNRS UMR 6002, Université Blaise Pascal and ENSCCF, 24 Avenue des Landais, Aubière 63177, France

[c] Centre d'Electronique et de Micro-optoélectronique de Montpellier, UMR 5507, Université de Montpellier II Place Eugène Bataillon, Montpellier cedex 5 34090, France

**Abstract**

New sol-gel derived organic/inorganic hybrid single mode waveguides devices have been developed for telecommunication applications in the two near infrared windows at 1310 and 1550 nm. The overall procedure of fabrication of these devices is described and the refractive indices of the guiding, the buffer and the protective layers are adjusted by a precise control of the materials' composition. Due to the improvement of the composition of the guiding layer, the attenuation losses are significantly decreased to 0.8 dB/cm and 2dB/cm at respectively 1310 and 1550 nm.

# 1. Introduction

In the past years, the expansion of telecommunication networks has induced the development of optical links using optical fibers. Integrated optical circuits are thought to solve various problems in routing, dividing, filtering the signal, and in reducing the number of electro-optic interconnections.

During the last few years, the fabrication of integrated optical devices using hybrid material has received an increasing amount of attention [1-5]. The technology that provides the lowest propagation losses is based on a two steps process [6]: fabrication of the guiding layer by the polymerization of a sol-gel precursor and imprinting of the channel waveguide by the photopolymerisation of an organic network using UV light through a mask. This technology allows a direct imprint of channel waveguides without etching and with no use of photoresist. Furthermore, high propagation losses induced by strong scattering effects due to the air cover and silicon substrate can easily be avoided by encapsulating the guide between a buffer layer and the composition of this protective layer can be adjusted in a way such that scratch resistance is enhanced to facilitate the handling of the circuit [6].

However, it is well known that both OH and CH groups present in hybrids materials compete with the propagation of the light by absorption processes involving their overtones and combinations [7]. The propagation losses measured in our reference material were as high as 1 dB/cm and 3.5 dB/cm at the second (1310 nm) and third (1550 nm) telecommunication windows respectively [8]. The low densification temperature treatment involved in standard hybrid sol-gel process leaves a too large OH amount, but unfortunately, this temperature cannot be increased due to the presence of the organic network. One efficient way to limit the amount of high vibrating molecular groups is to use mineral precursors that polymerize using a non-hydrolytic process [9], but this technique has at least two drawbacks: it is difficult to implement at an industrial scale because it requires a completely moisture free environment and it is not compatible with the above two-steps process.

Recently, the role of OH groups in the attenuation at 1550 nm was reconsidered, as it was shown that the absorption by CH aliphatic groups induced a stronger attenuation at this wavelength in organic/inorganic hybrids than OH groups [10,11]. Owing to these results, we have focused our attention on the preparation of a new organic/inorganic hybrid material containing a minimum amount of CH groups, using a reactive siloxane precursor associated with colloidal silica stabilized in a diacrylate organic precursor.

## 2. Experimental procedure

Buried channel waveguides were processed as a three-layer structure coated on a silicon substrate: a buffer layer (BL), a guiding layer (GL) and a protective layer (PL).

*2.1 The buffer and protective layers solution*

The role of the buffer and protective layers is to insulate the guide from the silicon substrate and from air and prevent from dust trapping. These two layers were prepared by the sol-gel synthesize of a copolymer made from methyl-triethoxysilane (MTEOS, Fluka, Assay > 98%) and phenyl-triethoxysilane (PHTEOS, Fluka, Assay > 98%). As the two alcoxides possess different reactivity, a three steps process was used, as described schematically in Fig. 1. The pre-hydrolysis of each alcoxide was first achieved separately using aqueous 0.01 N HCl solution with the same water to alcoxide ratio of 0.75:1, then the two sols were mixed without any precipitation. The last step consisted in a controlled hydrolysis of the resulting mixture in order to reach an equi-molar 1:1 water to alcoxide ratio.

*2.2 The active guiding layer solution*

The inorganic part was prepared as a tridimensionnal sol-gel processed structure using methyl-diethoxysilane (MDES, Fluka, Assay > 96%) as a siloxane precursor, with the aim to obtain a refractive index as close as possible to the one of the optical fibers and therefore minimize the coupling losses. For the organic part, we used the commercial precursor HIGHLINK-OG108 (from Clariant) which is composed of colloidal silica (particle of 25 nm size, 30 wt.%) stabilized in 70% of photosensitive tripropylene glycol diacrylate. The interest in the use of this precursor is to limit the amount of high vibrating CH groups carried by the diacrylate by dispersing these molecules in colloidal silica, which has no absorption at 1550 nm. After 30 minutes stirring, 2.5 wt.% of Irgacure 819 photoinitiator (from Ciba) was added to the MDES / HIGHLINK-OG108 mixture, as depicted in Fig. 1. No pre-hydrolysis was performed at this stage to allow a long time stability of the solution. Actually, the hydrolysis occurs during the deposition process under 50 % air moisture atmosphere.

*2.3 Fabrication of the waveguide*

The general procedure for the fabrication of waveguide involved the filtering of the above solutions through a 0.2 µm Teflon filter and the sequential deposition of the layers by the dip-coating technique, as sketched in Fig. 2. Care was taken to control appropriately the speed of the coatings in order to get a single layer thickness of 20 µm for BL and PL (this is to minimize scattering attenuation effects) and of 6 - 8 µm for GL. It should be noted that the cladding solution needed to be re-concentrated by extracting the solvent under vacuum prior to the coating. The GL was deposited in one step, under a controlled air moisture of 50 % and dried at 60 °C for 10 min to decrease mechanical resistance and allow direct imprint of the waveguide. The square section of the waveguide was typically 7x7 µm$^2$. The stabilization of the structure was completed by a post-back treatment at 120 °C for 2 hours. The ends of samples were cleaved for optical testing.

## 3. Results and discussion

To show the direct absorption at 1550 nm, a near infrared spectrum was recorded thorough a monolith prepared by gelation of the GL solution (Fig. 3). This spectrum is in perfect agreement with the ones of acrylic hybrid materials developed as photonic devices with the same technology [10]. As shown in previous papers, the absorption in the near infrared region consists of a superposition of several bands ascribed mainly to the organic part [10,11]. In this new composition containing a high amount of non absorbing colloidal silica, it is interesting to note that the absorption in the near infrared region is significantly reduced (by 20 % with respect to absorption in our reference material) so that we can expect a substantial improvement of the photonic performances at 1550 nm. Considering that guiding in such structures depends on the geometry and on the refractive index contrast between the guide and the cladding material (BL and GL), we determined the refractive indices of both GL and BL. In the reference material, single mode guiding was obtained for guides with a cross section of 7x7µm$^2$ and a refractive index contrast close to 12/1000. In our device, the refractive indices of the coatings were measured with an Abbe refractometer at 632.8 nm.

Figure 4 shows the variation of the refractive index of GL at 120°C (*viz.* the temperature of the final thermal stabilization of the optical device), as a function of the duration of the heat treatment. After one hour, the refractive index is 1.518. The value decreases to around 1.506 after two hours and then remains stable up to a four hours treatment. Considering that the refractive index reflects changes on the chemical composition and densification of the material, the above result tends to show that most of the physico-chemical modifications occur within the two first hours of heating. This behavior is different from the behavior of pure mineral materials that show an increase of the refractive index due to heat treatment densification [12]. In our hybrid material, the increase of the refractive index by densification is compensated by a predominant decrease of the molar refraction induced by polycondensation of silanols into siloxanes. This global decrease of the refractive index by sintering at a given temperature has already been observed for similar hybrid samples [13] and could constitute a specificity of this class of materials.

The refractive indices of the cladding material are plotted in Fig. 5 against the PHTEOS content for a heat treatment at 120°C for 4 hours. The refractive indices vary from 1.445 in pure PHTEOS to 1.515 for the highest concentrated materials. This increase of the refractive index is mainly due to the aromatic groups contained in PHTEOS which are known to possess a higher molar refraction than aliphatic groups. It is therefore interesting to note that a precise control of the PHTEOS content in the materials allows to adjust conveniently the refractive indices of the layers. In order to obtain a refractive index contrast close to 12/1000, a composition containing 15 % of PHTEOS was selected. A channel waveguide was made accordingly and was operated successfully at 630 nm, 1310 nm and 1550 nm, using a laser beam guided through an optical fiber. A near field picture of the output of the 7x7 $\mu m^2$ waveguide at 1550 nm is shown in Fig. 6. This waveguide was found to be single mode at 1310 and 1550 nm and the optical losses measured at these wavelengths were respectively 0.8 dB/cm and 2 dB/cm.

## 4. Conclusion

New single mode organic/inorganic hybrid waveguides operating at 1310 and 1550 nm have been processed by the photo-lithography technique. The fabrication of the guiding channel involved the sol-gel synthesize of aromatic and aliphatic precursors, with controlled refractive indices in the near infrared region. A diacrylate monomer associated with a compliant mineral network allowed to stabilized the coatings with a constant refractive index (1.502) after a heat treatment of two hours at 120 °C. Due to the improved composition of the waveguide, the attenuation is decreased by 20 % (viz. 0.8 dB/cm) at 1310 nm and around 45

% (viz. 2dB/cm) at 1550 nm with respect to our previous devices. These results confirm our previous work and are very encouraging for future development of optical devices with higher performances.

**Figure captions**

Figure 1 : Sol-gel synthesis of the cladding (BL and PL) and of the guiding (GL) materials.

Figure 2: Procedure for the preparation waveguide.

Figure 3: Near infrared spectrum of a monolith obtained after gelling of the GL sol.

Figure 4: Refractive index of the GL coating as a function of heat treatment time at 120°C.

Figure 5: Refractive index of the sheath material as a function of the PHTEOS content.

Figure 6: Near field picture of the output of a square structure guide (7x7µm$^2$) at 1550 nm.

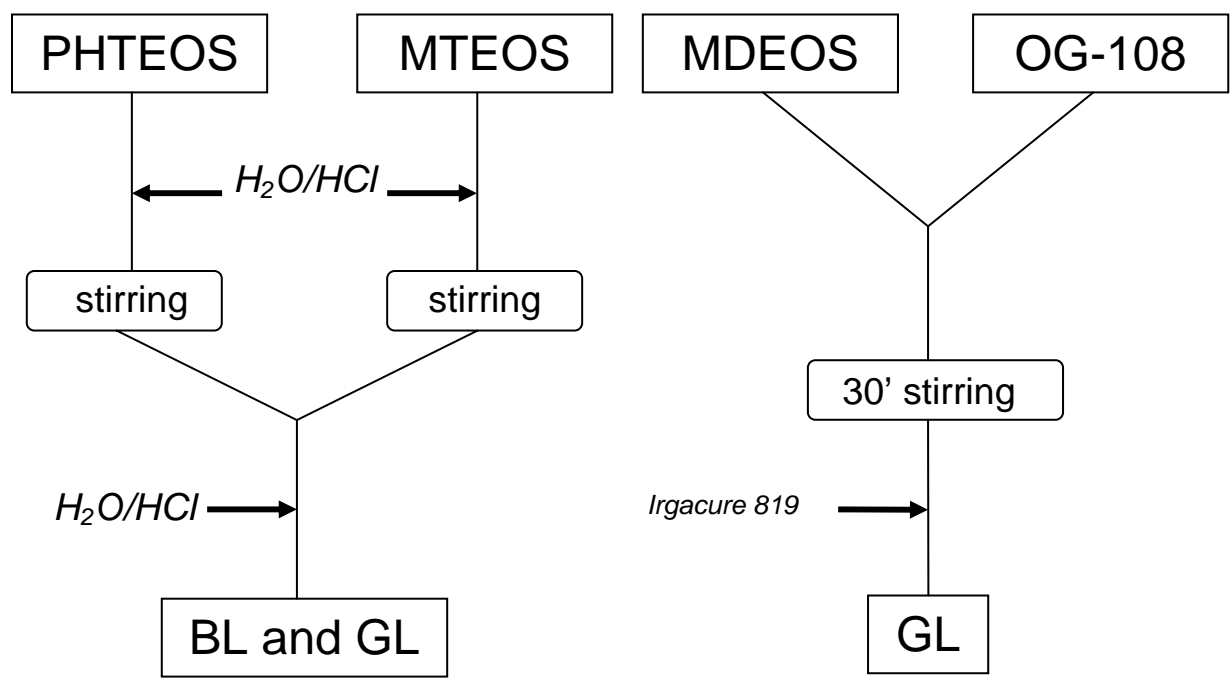

**Fig. 1**

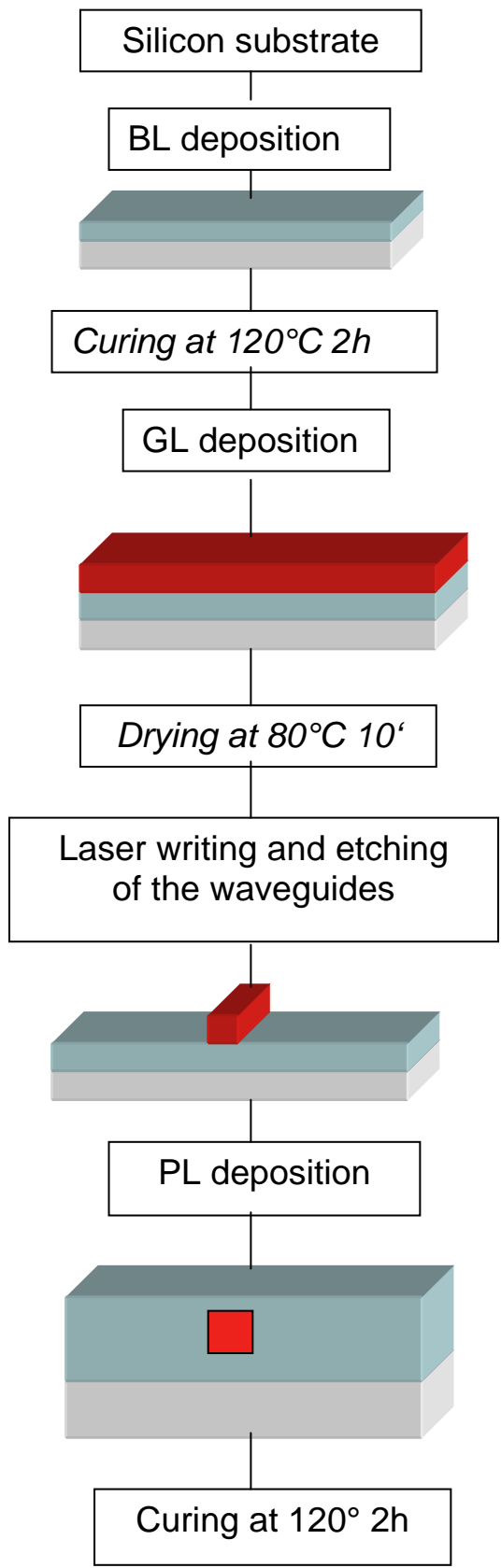

**Fig. 2**

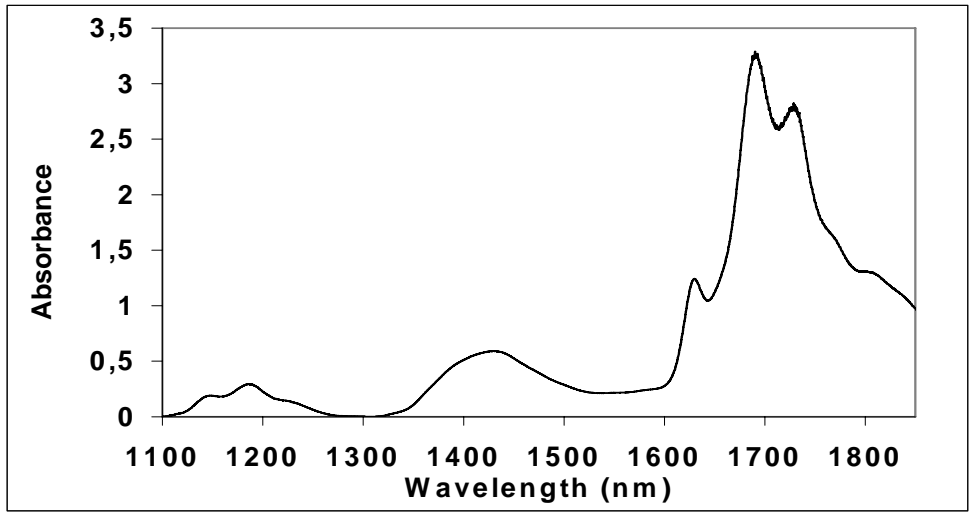

**Fig. 3**

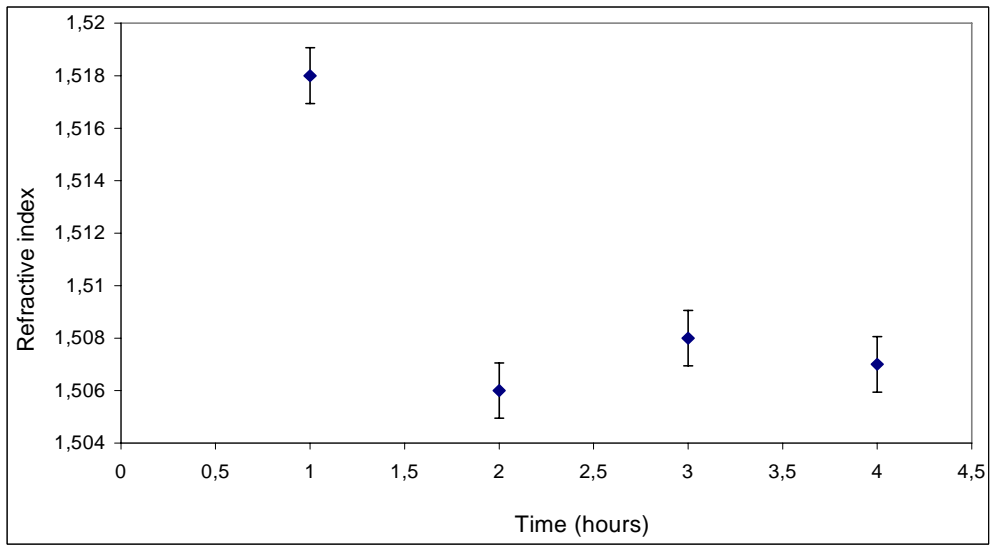

**Fig. 4**

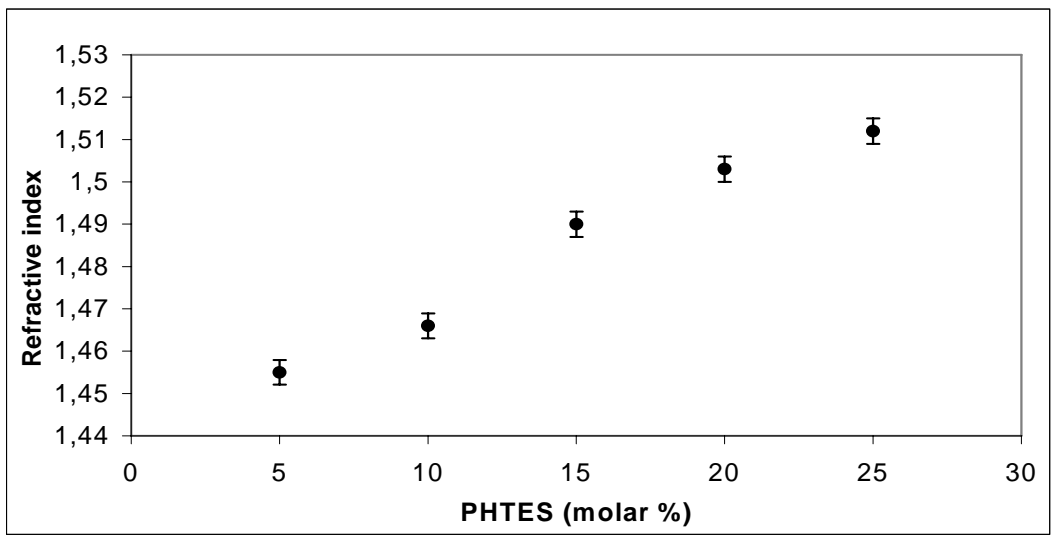

**Fig. 5**

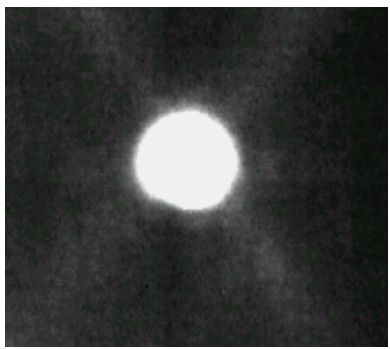

**Fig. 6**